\documentclass{icrc}

\usepackage{times}
\usepackage{graphicx} % when using Latex and dvips
\newcommand{\lwig}{\mbox{\,\raisebox{.3ex}
    {$<$}$\!\!\!\!\!$\raisebox{-.9ex}{$\sim$}\,}}

\begin{document}

\title{Possible 
detection of relic neutrinos and their mass
}
\author[1]{A. Ringwald}
\affil[2]{Deutsches Elektronen-Synchrotron DESY, Notkestr. 85, D-22607,
Hamburg, Germany}

\correspondence{A. Ringwald (ringwald@mail.desy.de)}

\firstpage{1}
\pubyear{2001}

% \titleheight{11cm} % uncomment and adjust in case your title block
                     % does not fit into the default and minimum 7.5 cm

\maketitle

\begin{abstract}
Recently the possibility was widely discussed that a large fraction of the 
highest energy cosmic rays may be decay products of Z bosons 
which were produced in the resonant annihilation of ultrahigh energy cosmic neutrinos
on cosmological relic neutrinos. If one takes this so-called Z-burst
scenario seriously, one may infer the mass of the heaviest relic neutrino 
as well as the necessary ultrahigh energy cosmic neutrino flux  
from a comparison of the predicted Z-burst spectrum
with the observed cosmic ray spectrum. 
\end{abstract}

\section{Introduction}

Big bang cosmology predicts the existence of a background gas of
free photons and neutrinos. The measured cosmic microwave background 
radiation (CMBR) supports the applicability of standard cosmology
back to photon decoupling which occured approximately one hundred thousand years
after the big bang. The relic neutrinos, on the other hand,
have decoupled when the universe had a temperature of $1$ MeV and an age
of just one second. Thus, a measurement of the relic neutrinos, with a
predicted average number density of 
\begin{equation}
\label{standard_number_dens}
\langle n_{\nu_i}\rangle = \langle n_{\bar\nu_i}\rangle
= \frac{3}{22}\, 
\underbrace{\langle n_{\gamma}\rangle}_{\rm CMBR}
\simeq 56\ {\rm cm}^{-3}\,,
\end{equation}
per light 
neutrino species $i$ ($m_{\nu_i}<1$ MeV), 
would provide a new window to the early universe. Their predicted number 
density is comparable to the one of the microwave photons. However,
since neutrinos interact only weakly, the relic neutrinos have not been detected until now.

\begin{figure}[hb]
\vspace*{2.0mm} % just in case for shifting the figure slightly down
\includegraphics[width=5.0cm]
{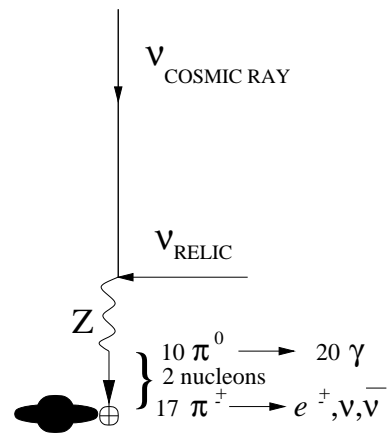}
\caption{\label{illu}
Illustration of a Z-burst resulting from the resonant annihilation of an 
ultrahigh energy cosmic neutrino on a relic (anti-)neutrino 
(adapted from Ref.~\citep{Pas:2001nd}).
}
\end{figure}

Recently, an indirect detection possibility for relic neutrinos has been discussed~~\citep{FMS99,W99}.
It is based on so-called Z-bursts resulting from the resonant annihilation of ultrahigh energy cosmic neutrinos
(UHEC$\nu$s) with relic neutrinos into $Z$ bo\-sons~\citep{W82,R93,Yoshida:1997ie}, the force carriers of the
electro\-weak neutral current (cf. Fig.~\ref{illu}). On resonance, the corresponding cross section is enhanced
by several orders of magnitudes. 
If neutrinos have non-vanishing masses $m_{\nu_i}$
-- for which there is ra\-ther convincing evidence in view of the observations of 
neutrino oscillations~\citep{Groom00} --
the resonance 
energies, in the rest system of the relic neutrinos, correspond to 
\begin{eqnarray}
\label{eres}
E_{\nu_i}^{\rm res} = \frac{M_Z^2}{2\,m_{\nu_i}} = 4.2\cdot 10^{21}\ {\rm eV}  
\left( \frac{1\ {\rm eV}}{m_{\nu_i}}\right)
\,,
\end{eqnarray}
with $M_Z$ denoting the mass of the Z boson. 
These resonance energies are, for neutrino masses of ${\mathcal O}(1)$ eV, remarkably
close to the energies of the highest energy cosmic rays. Indeed, it was argued~\citep{FMS99,W99}
that the ultrahigh energy cosmic rays (UHECRs) above the  
predicted Greisen-Zatsepin-Kuzmin (GZK) 
cutoff~\citep{G66,ZK66,BS00} ar\-ound $4\cdot 10^{19}$~eV
are mainly protons (and, maybe, photons) from Z decay. 

%%%%%%%%%%%%%%%%%%%%%%%%figure%%%%%%%%%%%%%%%%%%%%%%%%%%%%%%%%%%%%%%%
%%%%%%%%%%%%%%%%%%%%%%%%%%%%%%%%%%%%%%%%%%%%%%%%%%%%%%%%%%%%%%%%%%%%

This hypothesis was discussed in several 
papers~\citep{W98,Y99,GK99,Gelmini:2000ds,BVZ00,Pas:2001nd,FGDTK01,McKellar:2001hk}.
Here, we review the Z-burst scenario and report on a recent 
quantitative investigation~\citep{FKR01a,FKR01b}, where an attempt was made to
determine the mass of the heaviest relic neutrino as well as the 
necessary UHEC$\nu$ flux by comparing the predicted cosmic ray spectrum
from Z-bursts with the observed one via a maximum likelihood analysis.

\section{Z-burst spectrum}

A comparison of the Z-burst scenario with the observed ultrahigh energy
cosmic ray spectrum proceeds essentially in four steps. First, one has to determine the probability
of Z production as a function of the distance from Earth.
Secondly, one exploits data from collider experiments to derive 
the energy distribution of the produced protons in the lab system.
Thirdly, one has to take into account the propagation of the protons to Earth, i.\,e. one
has to determine their energy losses due to pion and $e^+e^-$ production 
through scattering on the CMBR and due to their redshift. The last
step is the comparison of the predicted and observed spectra and 
the extraction of the mass of the relic neutrino and the necessary UHEC$\nu$ flux.   

The contribution of protons 
from Z-bursts to the UHECR spectrum can be summarized as~\citep{FKR01a,FKR01b}
\begin{eqnarray}
\label{nu-flux}
\lefteqn{ F_{p|Z} ( E ) =  \sum_{i}
\int\limits_0^\infty {\rm d}E_p \int\limits_0^{R_0} {\rm d}r 
\int\limits_0^\infty {\rm d} \epsilon }  
\\[1ex] \nonumber && \times
\left[  F_{\nu_i}(E_{\nu_i},r)\, n_{\bar\nu_i}(r)
+  F_{\bar\nu_i}(E_{\bar\nu_i},r)\, n_{\nu_i}(r)
\right]
\\[1ex] \nonumber && \times
 \sigma_{\nu_i\bar\nu_i}( \epsilon)\,
{\rm Br}(Z\to {\rm hadrons})\,
{\cal Q}_{p+n}(E_p)
\\[1ex] \nonumber && \times
(-)\frac{\partial}{\partial E} P(r,E_p;E)
\,,
\end{eqnarray}
where $E$ is the energy of the protons arriving at Earth. 
Further important ingredients in Eq.~(\ref{nu-flux}) are: 
the ultrahigh energy cosmic neutrino fluxes $F_{\nu_i}(E_{\nu_i},r)$ at the resonant energies 
$E_{\nu_i}\approx E_{\nu_i}^{\rm res}$ and at distance $r$ to Earth, 
the number density 
$n_{\nu_i}(r)$ of the relic neutrinos, 
the Z production cross section $\sigma_{\nu_i\bar\nu_i}(\epsilon)$ at centre-of-mass (cm)
energy $\epsilon = \sqrt{2\,m_{\nu_i}\,E_{\nu_i}}$, the branching ratio 
for Z to hadrons, ${\rm Br}(Z\to {\rm hadrons})$, 
the energy distribution ${\cal Q}_{p+n}(E_p)$ of the produced protons 
(and neutrons) with energy $E_p$, and the probability $P(r,E_p;E)$ 
that a proton created at a distance $r$ with energy $E_p$ arrives
at Earth above the threshold energy $E$. 

The last four building blocks, $\sigma_{\nu_i\bar\nu_i}$, the hadronic branching ratio, 
${\cal Q}$, and $P$, are very well determined, whereas the  first two ingredients, 
the flux of UHEC$\nu$s, $F_{\nu_i}(E_{\nu_i},r)$, and the relic neutrino number 
density $n_{\nu_i}(r)$, are much less accurately known. 
In the following we shall discuss all these ingredients in detail.

\subsection{Z production and decay}
 
At LEP and SLC millions of Z bosons 
were produced and their decays analyzed with extreme high accuracy. 
Due to the large statistics, the uncertainties of the analysis  
related to Z decay are  negligible. 

\begin{figure}[t]
\vspace*{2.0mm} % just in case for shifting the figure slightly down
\includegraphics[bbllx=20pt,bblly=225pt,bburx=570pt,bbury=608pt,width=8.3cm]
{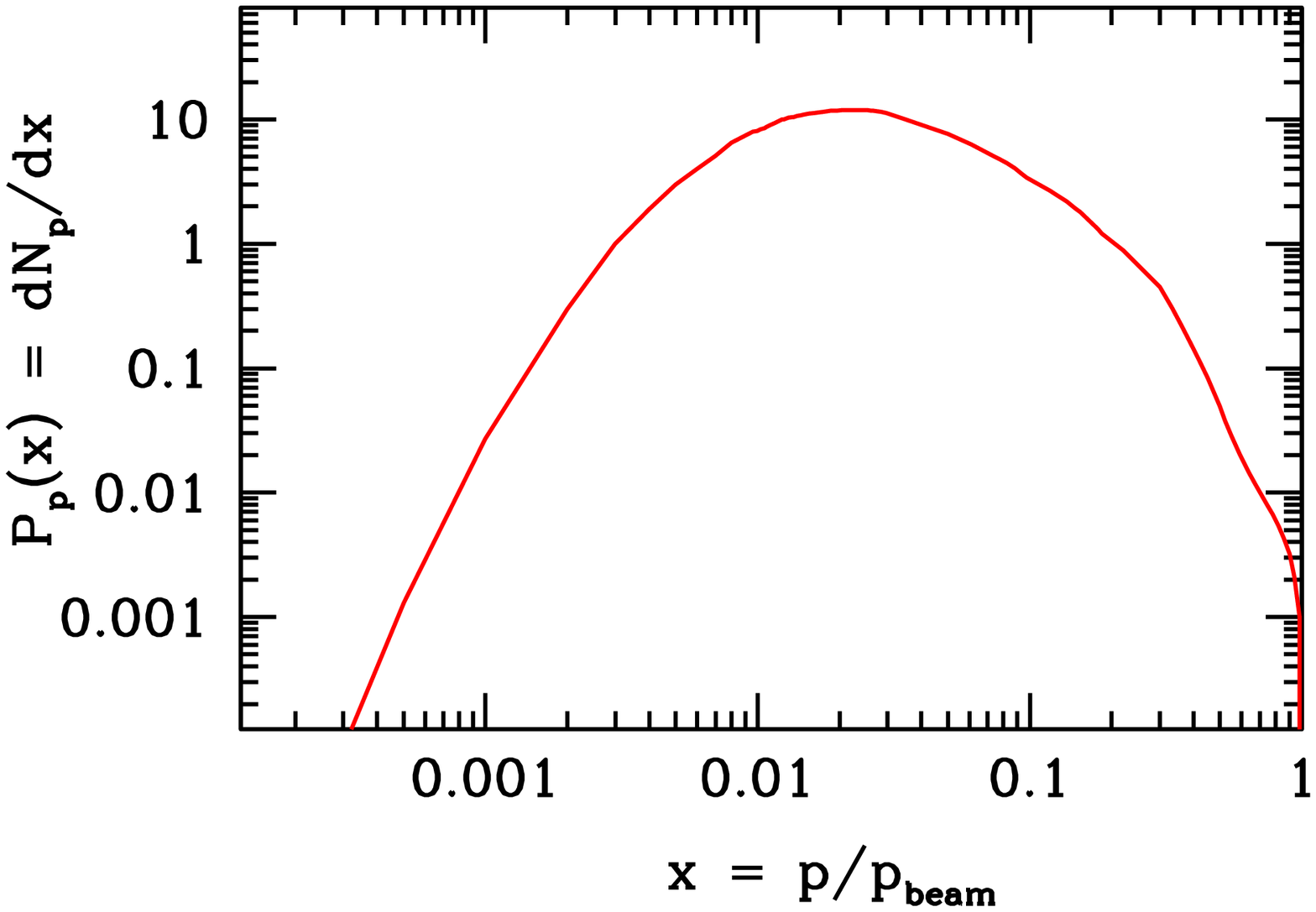}
\includegraphics[bbllx=20pt,bblly=225pt,bburx=570pt,bbury=608pt,width=8.3cm]
{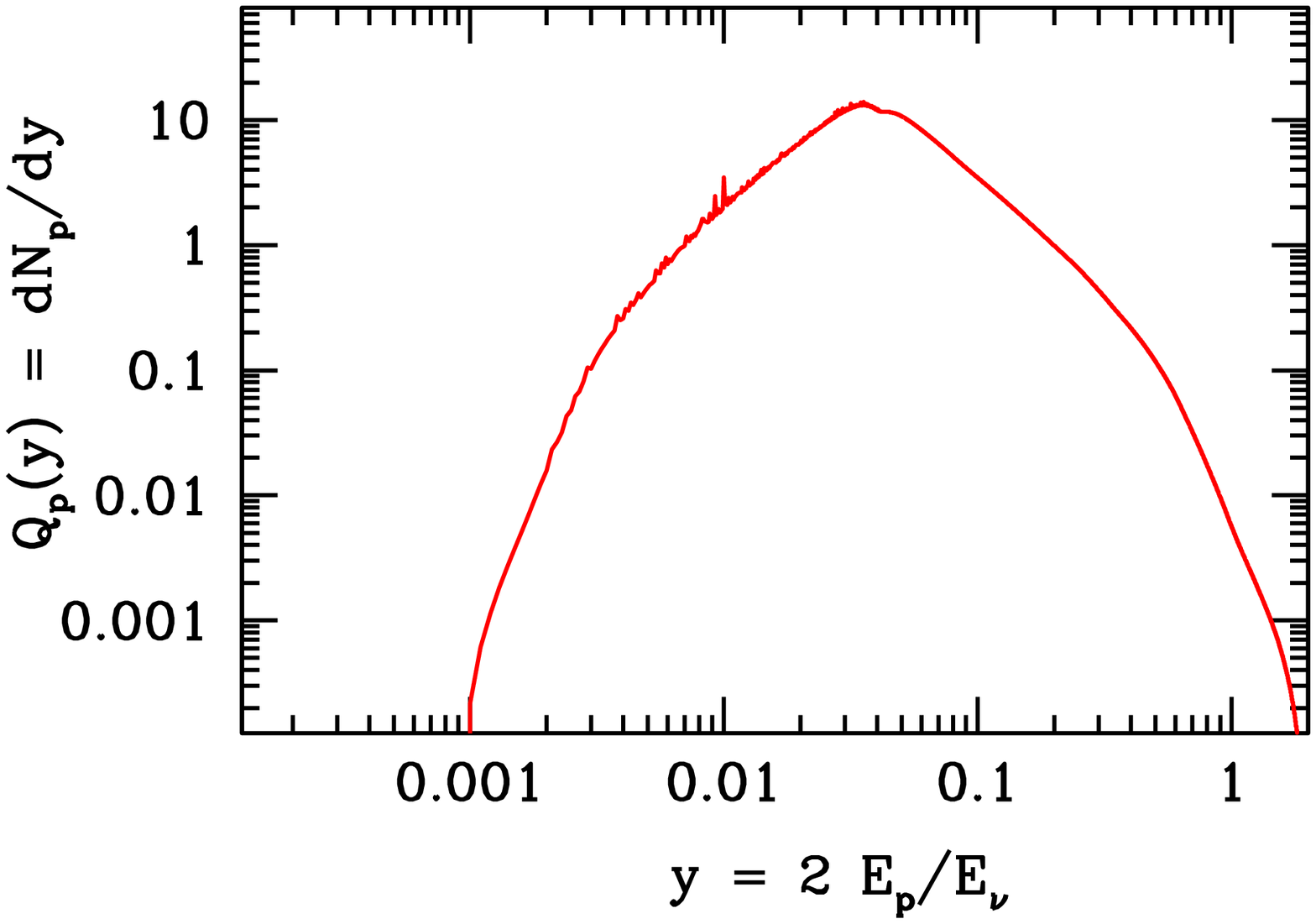} 
\caption{\label{prot-mom-dist}
Proton plus antiproton momentum distribution in Z decays (from Ref.~\citep{FKR01b}). 
{\em Top:} Combined data from Ref.~\citep{Akers94,Abreu95,Buskulic95,Abe99,Abe01}.
{\em Bottom:} Distribution in the lab system, in which the target 
relic neutrino is at rest. 
}
\end{figure}

\citet{FKR01a,FKR01b} 
combined 
existing published \citep{Akers94,Abreu95,Buskulic95,Abe99} and some improved 
unpublished~\citep{Abe01} data on the momentum 
distribution 
${\cal P}_p\,(x)={\rm d}N_p/{\rm d}x$, 
with momentum fraction $x=p_{\rm proton}/p_{\rm beam}$, 
of protons ($p$) 
(plus antiprotons ($\bar p$)) in Z 
decays, see Fig.~\ref{prot-mom-dist} (top). 
The $p+{\bar p}$ multiplicity is 
$\langle N_p\rangle = \int_0^1 {\rm d}x\, {\cal P}_p(x) = 1.04\pm 0.04$ in the 
hadronic channel~\citep{Groom00}. 

%%%%%%%%%%%%%%%%%%%%%%%%figure%%%%%%%%%%%%%%%%%%%%%%%%%%%%%%%%%%%%%%%
%%%%%%%%%%%%%%%%%%%%%%%%%%%%%%%%%%%%%%%%%%%%%%%%%%%%%%%%%%%%%%%%%%%%

The energy distribution ${\cal Q}_p(E_p)$ of the produced protons with energy 
$E_p$ entering in the spectrum~(\ref{nu-flux}) is obtained after a Lorentz transformation from the cm system
to the lab system, in which the target 
relic neutrino is at rest. It is only a function
of $y=2E_p/E_\nu$ and displayed in Fig.~\ref{prot-mom-dist} (bottom). 

Neutrons produced in Z decays will decay and end up as UHECR protons. 
They are taken into account according to 
\begin{equation}
{\cal Q}_{p+n}(y)=
\left(1+\frac{\langle N_n\rangle}{\langle N_p\rangle}\right)\,{\cal Q}_p(y)\,,
\end{equation}
where the neutron ($n$) (+ antineutron ($\bar n$)) multiplicity, 
$\langle N_n\rangle$,
is $\approx 4\%$ 
smaller than the proton's~\citep{Biebel01}. 

\subsection{Propagation of nucleons through the CMBR}

Similarly, the CMBR is known to a high accuracy.  It plays an important role in the 
determination~\citep{BW99,FK01a} of the probability $P(r,E_p,E)$, which takes 
into 
account the fact that protons of extragalactic (EG) origin and energies above 
$\approx 4\cdot 10^{19}$ eV 
lose a large fraction of their energies~\citep{G66,ZK66}
due to pion and $e^+e^-$ production 
through scattering on the CMBR and due to their redshift.
$P(r,E_p,E)$, in the form as it has been calculated for a wide range
of parameters by~\citet{FK01a}, is an 
indispensible tool in the quantitative analysis in Refs.~\citep{FKR01a,FKR01b}. 
In addition to its already published~\citep{FK01a} parametrized form, 
the numerical data for the probability distribution 
$(-)\partial P (r,E_p;E)/\partial E$ are now available via the 
World-Wide-Web URL {\it www.desy.de/\~{}uhecr}\,.

\subsection{UHEC\boldmath$\nu$\unboldmath\ fluxes}

%%%%%%%%%%%%%%%%%%%%%%%%figure%%%%%%%%%%%%%%%%%%%%%%%%%%%%%%%%%%%%%%%
\begin{figure}[t]
\vspace*{2.0mm} % just in case for shifting the figure slightly down
\includegraphics[bbllx=20pt,bblly=221pt,bburx=570pt,bbury=608pt,width=8.0cm]
{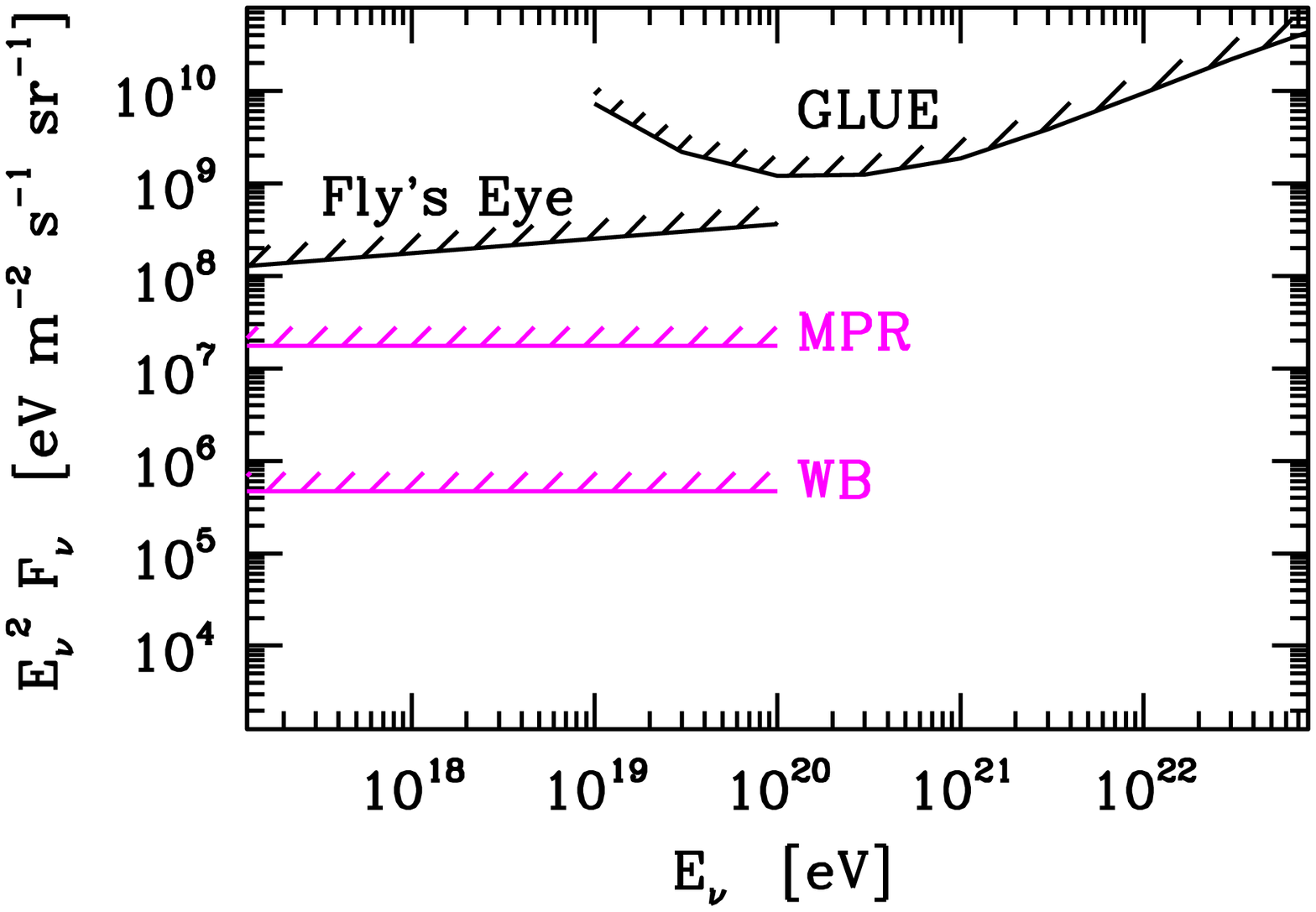}
\caption{\label{flux_upp_lim}
Upper limits on differential neutrino fluxes in the ultrahigh energy regime.
Shown are experimental upper limits 
on $F_{\nu_e}+F_{\bar\nu_e}$ from 
Fly's Eye~\citep{Baltrusaitis85} and 
on $\sum_{f=e,\mu} (F_{\nu_f}+F_{\bar\nu_f})$ from 
the Gold\-stone lunar ultrahigh energy neutrino ex\-pe\-ri\-ment 
GLUE~\citep{Gorham01}, as well as theoretical upper limits on 
$F_{\nu_\mu}+F_{\bar\nu_\mu}$
from ``visible'' (``WB''~\citep{Waxman:1999yy,Bahcall:2001yr}) and 
``hidden'' (``MPR''~\citep{Mannheim:2001wp}) hadronic astrophysical sources.
}
\end{figure}
%%%%%%%%%%%%%%%%%%%%%%%%%%%%%%%%%%%%%%%%%%%%%%%%%%%%%%%%%%%%%%%%%%%%

Presently unkown ingredients in the evaluation of the Z-burst spectrum~(\ref{nu-flux}) are 
the differential fluxes $F_{\nu_i}$ of ultrahigh energy cosmic neutrinos (see e.\,g.   
Refs.~\citep{Protheroe:1999ei,Gandhi:2000kq,Learned:2000sw} for recent reviews). 
Present experimental upper limits on these fluxes are rather poor 
 (cf. Fig.~\ref{flux_upp_lim} and contributions to these proceedings~\citep{Seckel:2001,Yoshida:2001}).

What are the theoretical expectations? 
More or less guaranteed are the so-called 
cosmogenic neutrinos which are produced when ultrahigh energy cosmic protons
scatter inelastically off the cosmic microwave background radiation~\citep{G66,ZK66}
in processes such as $p\gamma\to \Delta\to n\pi^+$,
where the produced pions subsequently 
decay~\citep{Beresinsky:1969qj,Beresinsky:1970}.
These fluxes (for recent estimates, see \citep{Yoshida:1993pt,Protheroe:1996ft,Yoshida:1997ie,Engel:2001hd}) 
represent reasonable lower limits on the 
ultrahigh energy cosmic neutrino flux, but turn out to be insufficient for the
Z-burst scenario. 
Recently, theoretical upper limits on the ultrahigh energy cosmic neutrino flux have been given 
by \citet{Waxman:1999yy}, \citet{Mannheim:2001wp}, and \citet{Bahcall:2001yr}. Per construction, the upper
limit from ``visible'' hadronic astrophysical sources, i.\,e. from those sources which are transparent to ultrahigh 
energy cosmic protons
and neutrons, is of the order of the cosmogenic neutrino flux~\citep{Waxman:1999yy,Mannheim:2001wp,Bahcall:2001yr}
and shown in Fig.~\ref{flux_upp_lim} (``WB''). Also shown in this figure (``MPR'') is 
the much larger upper limit from ``hidden'' hadronic astrophysical sources, i.\,e. from those sources from which only 
photons and neutrinos can escape~\citep{Berezinsky:1979pd,Mannheim:2001wp}.   

In this situation of insufficient knowledge, the following approach concerning the flux of 
ultrahigh energy cosmic neutrinos was taken in Refs.~\citep{FKR01a,FKR01b}.
The flux was assumed to have the form
\begin{equation} 
F_{\nu_i}(E_{\nu_i},r)=F_{\nu_i}(E_{\nu_i},0)\,(1+z)^\alpha\,,
\end{equation}
where $z$
is the redshift and where $\alpha$ characterizes the source evolution  
(see also~\citep{Yoshida:1997ie,Y99}). The flux at Earth, $F_{\nu_i}(E_{\nu_i},0)$,
or, more accurately, 
the sum of the corresponding fluxes at the resonant energies, 
\begin{equation}
F_Z = \sum_i \left[ F_{\nu_i}(E_{\nu_i}^{\rm res})+F_{\bar\nu_i}(E_{\nu_i}^{\rm res})
\right] \,,
\end{equation}
was then determined, together with the neutrino mass,  by a  fit to the UH\-E\-CR data.  
This analysis went up to distances $R_0$ (cf.~(\ref{nu-flux})) 
corresponding to redshift $z = 2$ (cf.~\citep{W95}), 
and uncertainties in the Hubble expansion rate $H_0=100\,h$ 
km\,s$^{-1}$\,{\rm Mpc}$^{-1}$, as given in Ref.~\citep{Groom00},
were included. 

\subsection{Relic neutrino number density}

%%%%%%%%%%%%%%%%%%%%%%%%figure%%%%%%%%%%%%%%%%%%%%%%%%%%%%%%%%%%%%%%%
\begin{figure}[t]
\vspace*{-5.0mm} % just in case for shifting the figure slightly down
\hspace*{-2.cm}
\includegraphics[angle=90,width=12cm]
{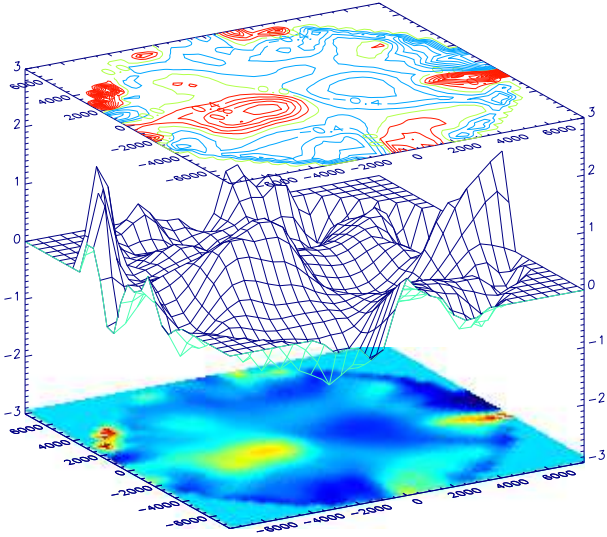}
\caption{\label{mass_dist}
Mass density fluctuation field $\delta%
% =(m_{\rm tot}-\langle m_{\rm tot}\rangle )/\langle m_{\rm tot}\rangle
$ along the 
supergalactic plane as obtained from
peculiar velocity measurements~\citep{daCosta:1996nt}. Shown are contours
in intervals of $\delta =2$, surface maps on a grid of spacing 
$500$ km\,s$^{-1}$, corresponding to $5\,h^{-1}$ Mpc, with the height proportional to
$\delta$, and contrast maps. One recognizes some well-known structures in the
nearby volume such as the Great Attractor at supergalactic coordinates
(SGX $\sim -2000$ km\,s$^{-1}$, SGY $\sim -500$ km\,s$^{-1}$),
the Perseus-Pisces complex 
(SGX $\sim 6000$ km\,s$^{-1}$, SGY $\sim -1000$ km\,s$^{-1}$),
and the large void
(SGX $\sim 2500$ km\,s$^{-1}$, SGY $\sim 0$ km\,s$^{-1}$) in between.
One can also see traces of the Coma/A1367 supercluster along 
SGY $\sim 7000$ km\,s$^{-1}$, and the Cetus region 
(SGX $\sim 500$ km\,s$^{-1}$, SGY $\sim -6000$ km\,s$^{-1}$).
}
\end{figure}
%%%%%%%%%%%%%%%%%%%%%%%%%%%%%%%%%%%%%%%%%%%%%%%%%%%%%%%%%%%%%%%%%%%%

The dependence of the relic neutrino number density $n_{\nu_i}$ on the distance $r$ 
was treated in the following way in Refs.~\citep{FKR01a,FKR01b}. 
The question is whether there is remarkable clustering of the relic neutrinos within
the local GZK zone of about 50 Mpc. It is known that the 
density distribution of relic neutrinos as hot dark matter follows the total mass 
distribution; however, with  less clustering~\citep{Ma:1998aw,Primack:2000iq}.
In fact, for $m_\nu\lwig 1$ eV, one expects pretty much that the neutrino number density
equals the big bang prediction~(\ref{standard_number_dens})~\citep{Primack:2001}. 
To take above facts into account, the shape of the $n_{\nu_i}(r)$ 
distribution was varied in Refs.~\citep{FKR01a,FKR01b}, 
for distances below 100 Mpc,  between the standard cosmological homogeneous case~(\ref{standard_number_dens}) 
and that of $m_{{\rm tot}}(r)$, the total mass distribution obtained from peculiar velocity 
measurements~\citep{daCosta:1996nt,Dekel99} (cf. Fig.~\ref{mass_dist}). 
These peculiar measurements suggest relative overdensities of at most a factor 2 $\div$ 3, 
depending on the grid spacing (cf. Fig.~\ref{mass_dist}). 
A relative overdensity of $10^2\div 10^4$ in our neighbourhood, as it was assumed in earlier 
investigations of the Z-burst hypothesis~\citep{FMS99,W99,W98,Y99,BVZ00}
(see also Ref.~\citep{McKellar:2001hk}), seems unnatural in view of these data. 
The quantitative results turned out to be rather insensitive to the 
variations of the overdensities within the considered range, whose effect were 
included in the  final error bars. For scales larger than 100 Mpc the relic neutrino 
density was taken according to the big-bang cosmology prediction, 
$n_{\nu_i}=56\cdot (1+z)^3$ cm$^{-3}$.

\section{Determination of $m_\nu$ and the UHEC$\nu$ flux}

\begin{figure}[t]
\vspace*{2.0mm} % just in case for shifting the figure slightly down
\includegraphics[bbllx=20pt,bblly=225pt,bburx=570pt,bbury=608pt,width=7.5cm]
{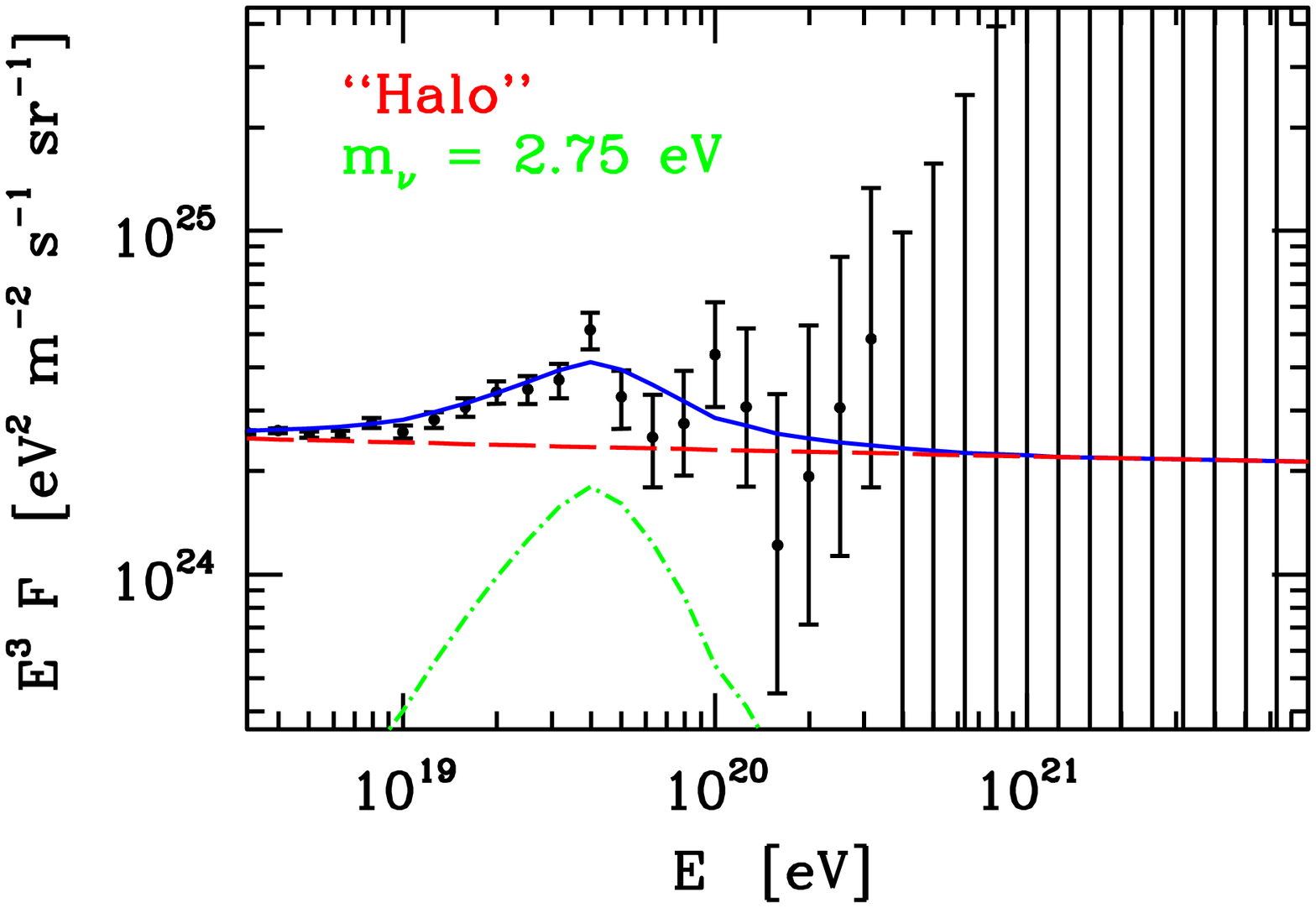}
\includegraphics[bbllx=20pt,bblly=225pt,bburx=570pt,bbury=608pt,width=7.5cm]
{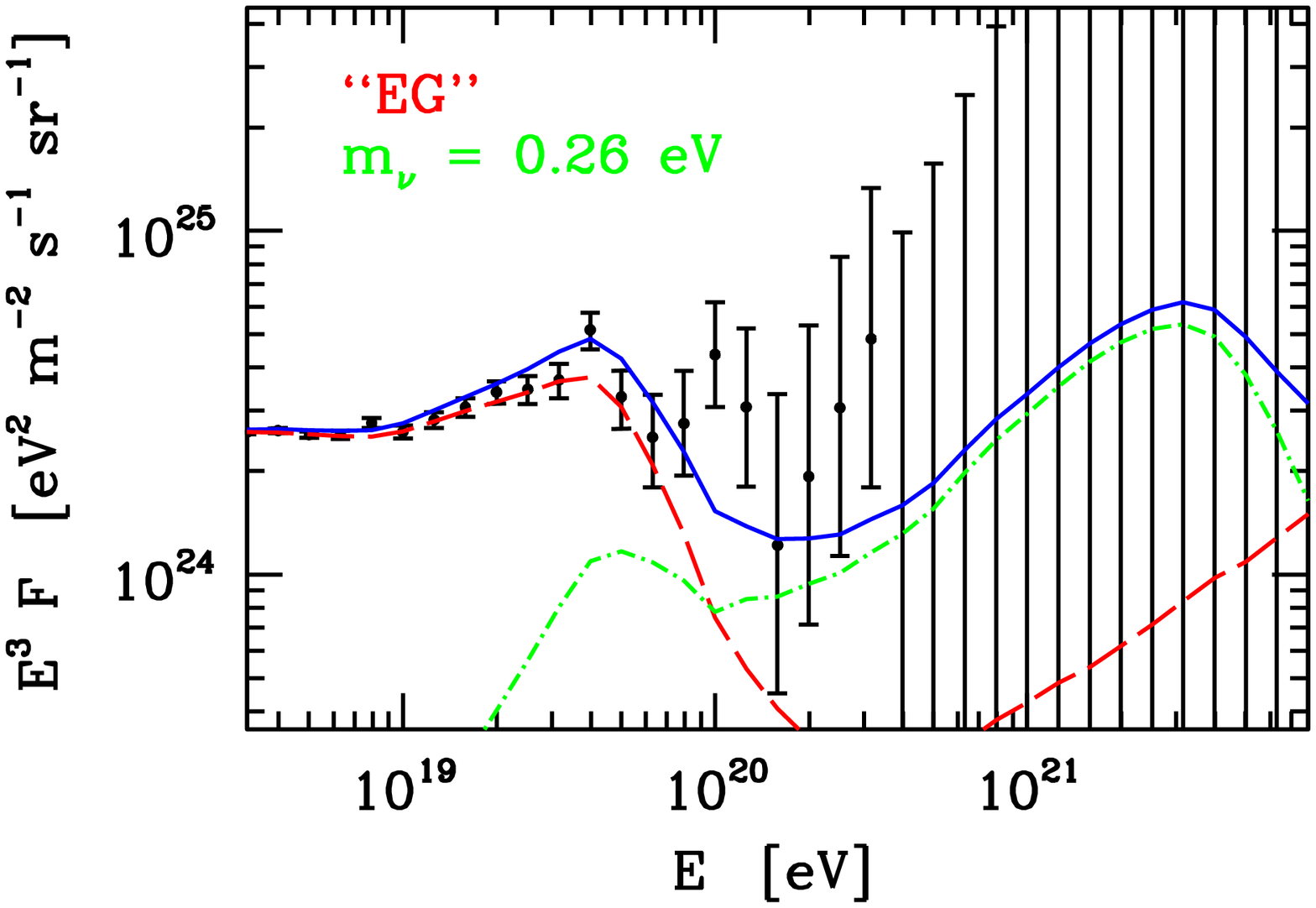} 
\caption{\label{fit_nu}
The available UHECR data with their error bars
and the best fits from Z-bursts, for nearly degenerate neutrino masses, $m_{\nu_i}\approx m_\nu$
and for cosmological evolution parameter 
$\alpha =3$~\citep{FKR01a,FKR01b}.
{\em Top:} Best fit for the ``halo''-case (solid line). The bump 
around $4\cdot 10^{19}$~eV is due to the Z-burst protons (dash-dotted), whereas the 
almost horizontal contribution (dashed) is the first, power-law-like 
term of Eq.~(\ref{flux}). {\em Bottom:} ``Extragalactic''-case.
The first bump at $4\cdot 10^{19}$~eV represents protons produced at
high energies and accumulated just above the GZK cutoff due to their energy
losses. The bump at $3\cdot 10^{21}$~eV is a remnant of the Z-burst
energy. The dashed line shows the contribution of the power-law-like
spectrum with the GZK effect included. The predicted fall-off for this
term around $4\cdot 10^{19}$~eV can be observed. 
}
\end{figure}

The predicted spectrum of protons from Z-bursts~(\ref{nu-flux})
can now be compared with the observed ultrahigh energy cosmic ray spectrum (cf. Fig.~\ref{fit_nu}). 
The analysis in Refs.~\citep{FKR01a,FKR01b} included 
published and unpublished (from the World-Wide-Web pages of the experiments on 17/03/01)
UHECR data
of AGA\-SA~\citep{Takeda98,Takeda99}, Fly's Eye~\citep{Bird93,Bird94,Bird95}, 
Ha\-ve\-rah Park~\citep{Lawrence91,Ave00}, 
and HIRES~\citep{Kieda00}. Due to normalization difficulties
the Ya\-kutsk~\citep{Efimov91} results were not used. 

As usual, each logarithmic unit between 
$\log (E/\mbox{eV})=18$ and $\log (E/\mbox{eV})=26$
was divided into ten bins. 
The predicted number of ultrahigh energy cosmic ray events in a bin was taken as
\begin{eqnarray}
\label{flux}
\lefteqn{ N(i)= }
\\[1ex] \nonumber &&
\int_{E_i}^{E_{i+1}}
{\rm d}E
\left[
F_{p|{\rm bkd}}(E;A,\beta )
+ F_{p|Z} (E; F_Z, m_{\nu_j} )\right],
\end{eqnarray}
where $E_i$ is the lower bound of the $i^{\rm th}$ energy bin. The first, 
background term $F_{p|{\rm bkd}}(E;A,\beta )$ was taken to have the usual 
power-law behavior which describes the data well for smaller 
energies~\citep{Takeda98,Takeda99}. 
The second term of the
flux in Eq.~(\ref{flux}) corresponds 
to the spectrum of the Z-bursts, Eq.~(\ref{nu-flux}). 

%%%%%%%%%%%%%%%%%%%%%%%%figure%%%%%%%%%%%%%%%%%%%%%%%%%%%%%%%%%%%%%%%
%%%%%%%%%%%%%%%%%%%%%%%%%%%%%%%%%%%%%%%%%%%%%%%%%%%%%%%%%%%%%%%%%%%%

Two possibilities for the background term were studied. 
In the first case it was assumed that the power part is produced in our galaxy. Thus no
GZK effect was included for it (``halo''), and it was taken as,
\begin{equation}
\label{pow-law-halo}
F_{p|{\rm bkd}}(E;A,\beta ) = A \cdot E^{-\beta}
 \hspace{6ex} ({\rm halo})\,.
\end{equation}
In the second -- in
some sense more realistic -- case 
it was assumed that the background protons come from uniformly distributed, extragalactic sources
and suffer from the GZK cutoff (``EG''). In this case, 
$A \cdot E^{-\beta}_p$ was taken as an injection spectrum, and the simple
power-law-like term was modified, by taking into account the probability 
$P(r,E_p;E)$, 
\begin{eqnarray}
\label{pow-law-eg}
\lefteqn{
F_{p|{\rm bkd}}(E;A,\beta ) = 
 \int_0^\infty {\rm d}E_p \int_0^{R_0} {\rm d}r }
\\[1ex]\nonumber && 
\times\,
A \cdot E^{-\beta}_p\,
(-)\frac{\partial P(r,E_p;E)}{\partial E}
 \hspace{6ex} ({\rm EG})\,.
\end{eqnarray}
It falls off around $4\cdot 10^{19}$~eV (see Fig.~\ref{fit_nu} (bottom)). 

Note that the following implicit assumptions have been made through 
ansatz~(\ref{flux}): 
{\em i)} It was assumed that the ultrahigh energy photons from 
Z-bursts can be neglected. 
{\em ii)} It was assumed that there are no significant additional primary 
ultrahigh energy  proton fluxes beyond the extrapolation of the above power-law. 
Assumption {\em i)} can be justified on account of the result of a 
detailed study~\citep{FKR01b} of the boosted Z-decay 
(data from Refs.~\citep{Akers94,Abreu95,Buskulic95,Abe99,Abe01}) which show that 
the energy of the photons is peaked at $1.7\cdot 10^{18}$ (1 eV/$m_{\nu_i}$) eV, 
at which the attenuation length of photons is much smaller than that of protons. 
Thus, the contribution of ultrahigh energy photons from Z decay in the observed 
UHECR spectrum is far less relevant than that of the protons. 
Assumption {\em ii)} will be relaxed in Ref.~\citep{FKR01b}.

The expectation value for the number of events in a bin is given
by Eq.~(\ref{flux}). 
To determine the most probable value for $m_\nu$ the maximum 
likelihood method was used and the 
$\chi^2(\beta,A,F_Z,m_\nu)$~\citep{FK01b},
\begin{equation} \label{chi}
\chi^2=\sum_{i=18.5}^{26.0}
2\left[ N(i)-N_{\rm o}(i)+N_{\rm o}(i)
\ln\left( N_{\rm o}(i)/N(i)\right) \right],
\end{equation}
was minimized, 
where $N_{\rm o}(i)$ is the total number of observed events in the $i^{\rm th}$
bin. Since the Z-burst scenario results
in a quite small flux for lower energies, the ``ankle''  
was used as a lower end for the UHECR spectrum:
$\log (E_{\rm min}/\mbox{eV})=18.5$. The results are
insensitive to the definition of the upper end ( -- the flux is
extremely small there -- ) for which $\log (E_{\rm max}/\mbox{eV})=26$ was chosen.
The uncertainties of the
measured energies are about 30\% which is one bin. Using a Monte-Carlo me\-thod
this uncertainty was included in the final error estimates.

In the case where there are three neutrino types with
nearly degenerate neutrino masses, $m_{\nu_i}\approx m_\nu$,  
the fitting procedure involves four parameters: $\beta,A,F_Z$ and 
$m_\nu$. The minimum of the $\chi^2(\beta,A,F_Z,m_\nu)$ function is 
$\chi^2_{\rm min}$ at $m_{\nu\, {\rm min}}$ which is
the most probable value for the mass, whereas
the 1\,$\sigma$ (68\%) confidence interval for $m_\nu$ 
is determined by 
\begin{equation}
\chi^2(\beta',A',F_Z',m_\nu)\equiv \chi^2_o(m_\nu)=\chi^2_{\rm min}+1
\,.
\end{equation}
Here $\beta'$, $A'$, $F_Z'$ are defined in such a way that the 
$\chi^2$ function is minimized in $\beta,A$
and $F_Z$, at fixed $m_\nu$.

Qualitatively, the analysis can be understood in the following way.
In the Z-burst scenario a small relic neutrino mass needs large 
$E_\nu^{\rm res}$ in order to produce a Z. Large $E_\nu^{\rm res}$ results in 
a large Lorentz boost, thus large $E_p$. In this way the {\em shape} of the
detected energy ($E$) spectrum determines the mass of the relic neutrino. 
The sum of the necessary UHEC$\nu$ fluxes  $F_Z$, on the other hand, is determined by the 
over-all {\em normalization}.

The best fits to the observed data can be seen in 
Fig.~\ref{fit_nu}, for evolution parameter $\alpha=3$. 
For the ``halo''-case, a neutrino mass of
$2.75^{+1.28(3.15)}_{-0.97(1.89)}$~eV was found, whereas,  
for the ``EG''-case, the fit yielded $0.26^{+0.20(0.50)}_{-0.14(0.22)}$~eV.
The first numbers are
the 1\,$\sigma$, the numbers in the brackets are the
2\,$\sigma$ errors. This gives an absolute lower bound on the mass of the 
heaviest $\nu$ of $0.06$ eV at the 95\% confidence level (CL). 
Note, that the surprisingly small uncertainties are based on the above
$\chi^2$ analysis and dominantly statistical ones.
The fits are
rather good; for 21 non-vanishing bins and 4 fitted parameters
they can be as low as 
$\chi^2=18.6$. 
The neutrino mass was determined for a wide range
of cosmological source evolution ($\alpha =0\div 3$) and Hubble 
parameter 
($H_0=(71\pm 7)\times^{1.15}_{0.95}$~km/sec/Mpc) and was found to depend only moderately  
on them. The results remained within the above error bars. 
A Monte-Carlo analysis was performed studying higher statistics. In the near 
future, Auger~\citep{Guerard99,Bertou00} will provide a ten times 
higher statistics, which 
reduces the error bars in the neutrino mass to $\approx$ one third of their 
present values.  

%%%%%%%%%%%%%%%%%%%%%%%%figure%%%%%%%%%%%%%%%%%%%%%%%%%%%%%%%%%%%%%%%
\begin{figure}[t]
\vspace*{2.0mm} % just in case for shifting the figure slightly down
\includegraphics[bbllx=20pt,bblly=221pt,bburx=570pt,bbury=608pt,width=8.0cm]
{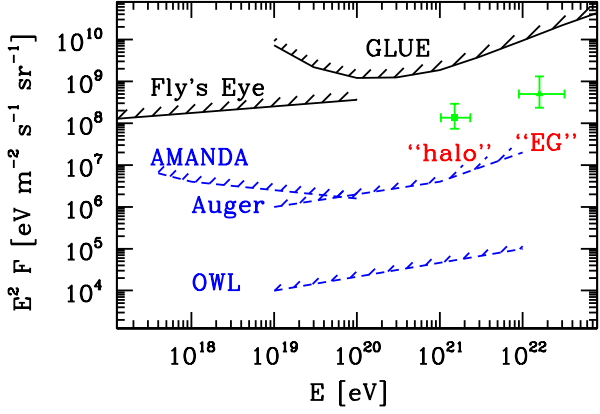}
\caption{\label{eflux}
Neutrino flux, $F = \frac{1}{3}\sum_{i=1}^3 ( F_{\nu_i}+F_{\bar\nu_i})$, 
required by the Z-burst hypothesis for 
the ``halo'' and the ``extragalactic'' case, 
for evolution parameter $\alpha =3$~\citep{FKR01a,FKR01b}.
The horizontal errors indicate the 1\,$\sigma$ uncertainty of the
mass determination and the vertical errors include also the uncertainty
of the Hubble expansion rate.
The dependence on 
$\alpha$ is just of the order of the thickness of the lines.
Also shown are upper limits from Fly's Eye~\citep{Baltrusaitis85} and the 
Gold\-stone lunar ultrahigh energy neutrino ex\-pe\-ri\-ment 
GLUE~\citep{Gorham01}, 
as well as projected sen\-si\-tivi\-ties of AMAN\-DA~\citep{Barwick00}, 
Auger~\citep{Y99,Capelle98} and OWL~\citep{Y99,Ormes97}.}
\end{figure}
%%%%%%%%%%%%%%%%%%%%%%%%%%%%%%%%%%%%%%%%%%%%%%%%%%%%%%%%%%%%%%%%%%%%

\begin{figure}[t]
\vspace*{2.0mm} % just in case for shifting the figure slightly down
\includegraphics[width=8.4cm]
{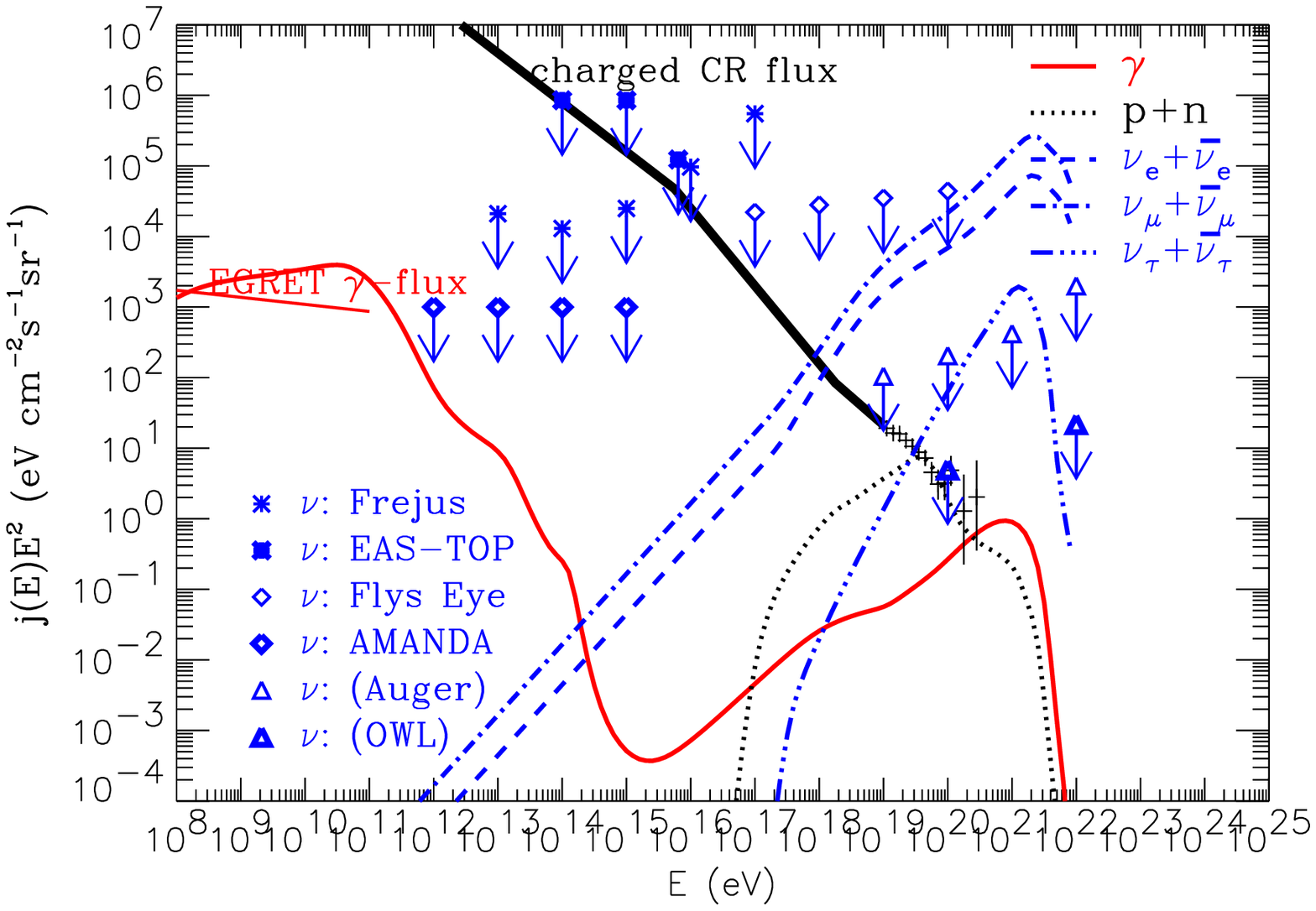}
\includegraphics[width=8.4cm]
{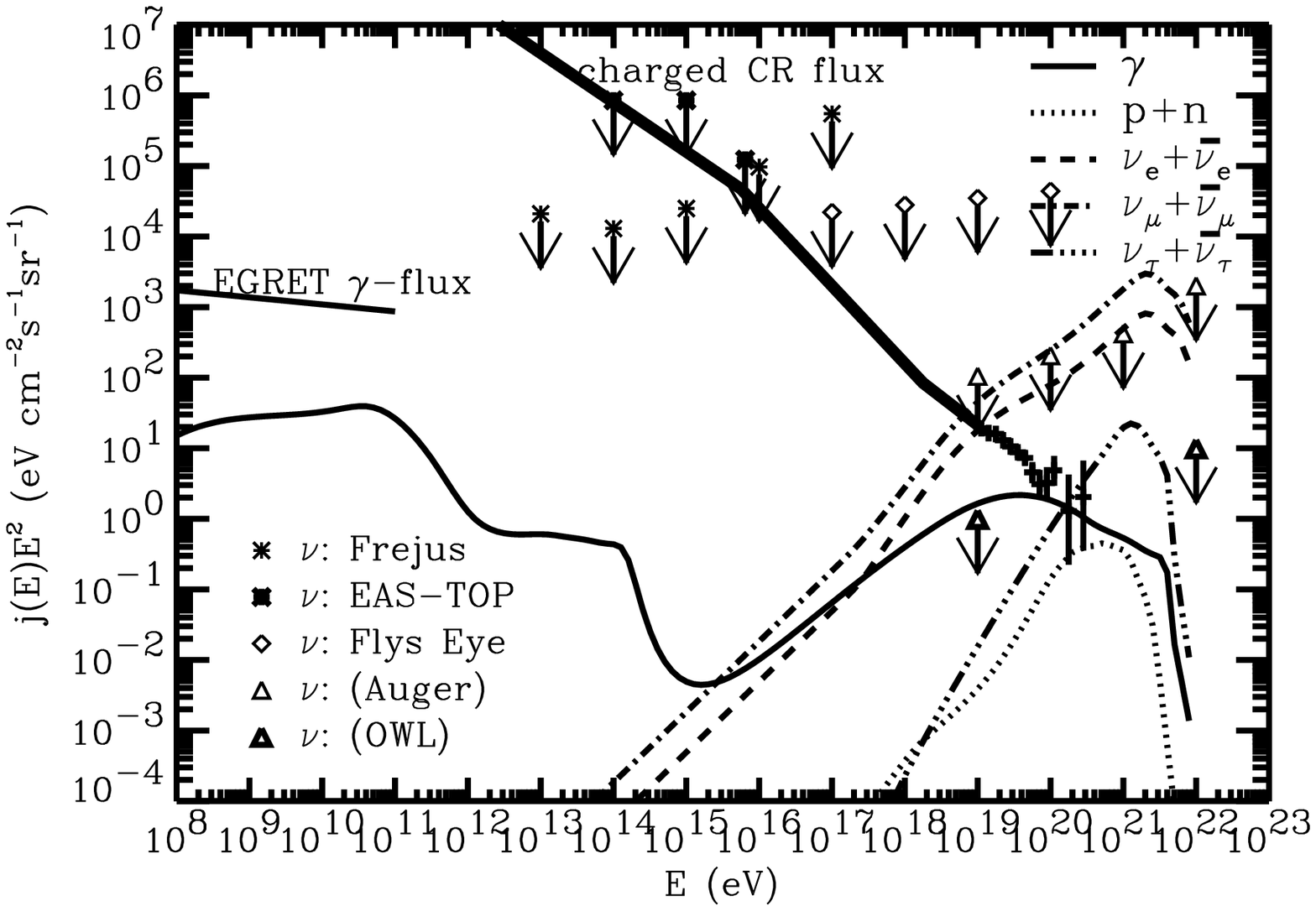} 
\caption{\label{comp_ysl}
Spectra of nucleons, $\gamma$ rays, and neutrinos for three degenerate
neutrinos $m_{\nu_i}= m_\nu =1$ eV, for 
cosmological evolution parameter $\alpha =3$ and an input differential
spectrum of ultrahigh energy cosmic neutrinos $\propto E_\nu^{-1}$.
{\em Top:} Obtained with the standard cosmological value~(\ref{standard_number_dens}) 
for the relic neutrino number 
density~\citep{Sigl:priv}.
{\em Bottom:} Obtained with a local overdensity of $300$ on a scale of 5 Mpc over the 
the standard cosmological value~(\ref{standard_number_dens}) for the relic neutrino number density~\citep{Y99}.
}
\end{figure}

The necessary UHEC$\nu$ fluxes at $E_\nu^{\rm res}$ have been obtained from the fits. 
They are summarized in Fig.~\ref{eflux}, together with some
existing upper limits and projected sensitivities of 
pre\-sent, near future and future observational projects. 
The necessary UHEC$\nu$ flux appears to be well below present upper limits
and is within the expected sensitivity of AMANDA, Auger, and OWL.
An important constraint for all top-down scenarios~\citep{BS00} is the 
EGRET observation of a diffuse $\gamma$ background~\citep{Sreekumar}.   
As a cross check, one may calculate the total energy in photons from Z-bursts. 
Even if one assumes that all their energy ends up between 30 MeV and 100 GeV, one
finds that the $\gamma$ flux is somewhat smaller than that of EGRET.
These numerical findings are in fairly good agreement with simulations done by~\citet{Sigl:priv} 
(cf. Fig.~\ref{comp_ysl} (top)). 

\section{Comparison with $\triangle m_\nu^2$ from neutrino oscillations}

%%%%%%%%%%%%%%%%%%%%%%%%figure%%%%%%%%%%%%%%%%%%%%%%%%%%%%%%%%%%%%%%%
%%%%%%%%%%%%%%%%%%%%%%%%%%%%%%%%%%%%%%%%%%%%%%%%%%%%%%%%%%%%%%%%%%%%

One of the most attractive patterns for neutrino masses is si\-milar to
the one of the charged leptons or quarks: the masses are 
hierarchical, thus the mass difference between the families is 
approximately the mass of the heavier particle. Using the 
mass difference of the atmospheric neutrino oscillation for the
heaviest mass~\citep{Groom00}, one obtains values between 0.03 and 0.09~eV.  
It is an intriguing feature of the result found in Refs.~\citep{FKR01a,FKR01b}
that the smaller one of the predicted masses is compatible on the 
$\approx$ 1.3\,$\sigma$ level with this scenario. 

\begin{figure}[t]
\vspace*{2.0mm} % just in case for shifting the figure slightly down
\includegraphics[bbllx=20pt,bblly=221pt,bburx=570pt,bbury=608pt,width=8.3cm]
{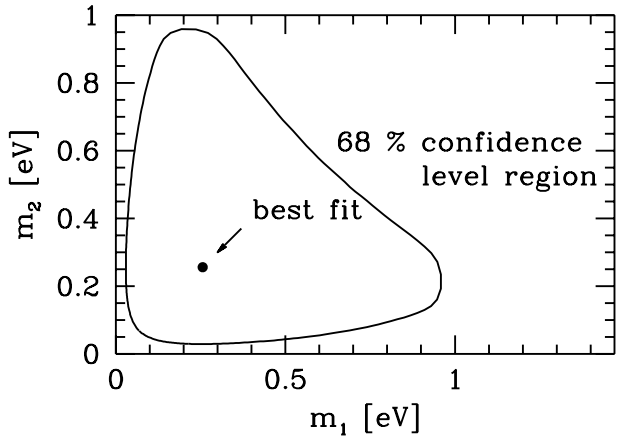}
\caption{\label{masses}
{
The best fit and the 1\,$\sigma$ (68\% confidence level) region in a scenario
with two non-degenerate neutrino masses~\citep{FKR01a,FKR01b}.
}}
\end{figure}

Another popular possibility is to have 4 neutrino types. Two of them 
-- electron and sterile neutrinos -- are
separated by the solar neutrino oscillation solution, the other two
-- muon and tau -- by the atmospheric neutrino oscillation 
solution, whereas the mass difference between the two groups is
of the order of 1~eV. This possibility was studied, too. On the 
mass scales and resolution considered, the electron and sterile neutrinos are 
practically degenerate with mass $m_1$ and the muon and tau
neutrinos are also degenerate with mass $m_2$. The best fit and the  
1\,$\sigma$ region in the $m_1-m_2$ plane  
is shown in Fig.~\ref{masses} for the ``EG''-case for the background protons. 
The dependence of this result on the cosmological evolution and 
on the UHEC$\nu$ spectrum will be discussed in Ref.~\citep{FKR01b}.
Since
this two-mass scenario has much less constraints the allowed region 
for the masses is larger than in the one-mass scenario.   

%%%%%%%%%%%%%%%%%%%figure%%%%%%%%%%%%%%%%%%%%%%%%%%%%%%%%
%%%%%%%%%%%%%%%%%%%%%%%%%%%%%%%%%%%%%%%%%%%%%%%%%%%%%%%%%

\section{Comparison with other studies}

Numerical simulations of Z-burst cascades for $m_\nu\sim 1$ eV, 
taking into account all known extragalactic propagation effects, were performed
by \citet{Y99}. Based on case studies, local overdensities by factors 
ranging from $20\div 10^3$ over the standard cosmological relic neutrino number density 
on a scale of 5 Mpc were argued to be necessary in order to 
get a successful description of the UHECR events and rate above the
GZK cutoff without violating lower energy photon flux limits and
without invoking inconceivable UHEC$\nu$ fluxes. For such large overdensities,
most of the UHECRs from Z-bursts originate nearby and their 
attenuation to the Earth can be neglected. In our case, with realistic
overdensities $\lwig 2\div 3$ on scales $\lwig 100$ Mpc, most of 
the UHECRs from Z-bursts originate from cosmological distances and suffer therefore
much more from GZK attenuation effects. Therefore, 
despite of the fact that by 
construction the overall rate of UHECRs from Z-bursts observed at Earth 
is the same in both investigations, the predicted spectra 
are quite different, as illustrated in Fig.~\ref{comp_ysl}. 
For example, in the case of a large neutrino overdensity, 
the Z-burst proton spectrum has a single bump, reflecting that of 
Fig.~\ref{prot-mom-dist} (bottom), at some fraction of
the resonance energy (cf. Fig.~\ref{comp_ysl} (bottom)),
whereas it is seriously deformed in the case with a realistic relic neutrino
number density, piling up near the GZK cutoff (cf. Figs.~\ref{fit_nu} and \ref{comp_ysl} (top)).
Moreover, whereas, in the overdense case, the contribution of Z-burst photons dominates, in the presently 
observed ultrahigh energy region, over the one of Z-burst 
protons~(cf. Fig.~\ref{comp_ysl} (bottom)),
it can be neglected in the non-overdense case~(cf. Fig.~\ref{comp_ysl} (top)).   

Both variants of the Z-burst scenario, the one with and the one without 
remarkable local clustering of relic neutrinos, have problems. In the former
case one has to admit that no mechanism is known which leads to such 
a large overdensity. Gravitational clustering is really too inefficient for neutrinos with 
$m_\nu \lwig {\mathcal O}(1)$ eV~\citep{Ma:1998aw,Primack:2000iq,Primack:2001}. 
The case without remarkable local overdensity, on the other hand, requires 
a tremendous ultrahigh energy cosmic neutrino flux (cf. Fig.~\ref{eflux}), which, however, 
may be mildered by a factor of ${\mathcal O}(3)$ if one invokes, 
in extension of the standard cosmology, lepton asymmetries~\citep{GK99} at a level which 
is still allowed~\citep{Kneller:2001cd}. 
The required neutrino flux is larger than the theoretical upper limit from ``hidden'' hadronic 
astrophysical sources, from which only 
photons and neutrinos can escape (cf. Fig.~\ref{flux_upp_lim} (``MPR'')).
From an analysis of the assumptions behind this limit~\citep{Mannheim:2001wp} 
one finds that one has to invoke hidden astrophysical sources whose photons somehow
don't show up in the diffuse $\gamma$ ray background measured by EGRET~\citep{Sreekumar}.
One of the possibilities are sources whose surrounding is so dense that no 
ultrahigh energy photons can escape. 
It is an interesting question whether such conditions can be realized in BL Lacertae objects, a 
class of active galactic nuclei recently discussed as possible sources of the
highest energy cosmic rays~\citep{Tinyakov:2001nr}.
Alternatively, one may invoke top-down scenarios~\citep{BS00} for
the sources of the highest energy cosmic neutrinos such as unstable superheavy relic
particle decays~\citep{Gelmini:2000ds}.        

\section{Conclusions}

I reviewed a recent comparison of the predicted spec\-trum from 
Z-bu\-rsts -- resulting from the resonant annihilation of ultrahigh energy cosmic neutrinos
with relic neutrinos --  
with the observed
ultrahigh energy cosmic ray spectrum~\citep{FKR01a,FKR01b}. 
The mass of the heaviest relic neutrino turned out to be 
$m_\nu=2.75^{+1.28}_{-0.97}$ eV 
for halo- and
$0.26^{+0.20}_{-0.14}$~eV for extragalactic-scenarios for the background protons. 
The second mass, 
with a lower bound of $0.06$ eV on the 95\% CL, is 
compatible with a 
hierarchical neutrino mass scenario with the largest mass suggested by 
the atmospheric neutrino oscillation. 
The above neutrino masses are in the range 
which can be explored by future laboratory experiments
like the $\beta$ decay endpoint spectrum  and the 
neutrinoless $\beta\beta$ 
decay~\citep{Pas:2001nd,Vissani:2001ci,Farzan:2001cj,Osland:2001pm,Czakon:2001uh}
or by simultaneous observations of the bursts of neutrinos and gravitational
waves emitted during a stellar collapse~\citep{Arnaud:2001gt}.
They compare also favourably with the upper limit  
$\sum_i m_{\nu_i}\leq 4.4$ eV found recently from a global cosmological analysis involving also the 
recent CMBR measurements~\citep{WTZ01}, with a future sensitivity down to $\lwig 0.3$ eV~\citep{Hu:1998mj}.  
We analysed a possible two-mass scenario and gave the corresponding confidence
level region. The necessary UHEC$\nu$
flux was found to be consistent with present upper limits
and detectable in the near future. Astrophysical sources of these ultrahigh energy cosmic
neutrinos should be hidden even in $\gamma$ rays. 

\begin{acknowledgements}
I thank Z. Fodor and S. Katz for the nice collaboration and for a careful
reading of the manuscript. Furthermore, I thank S. Barwick, 
O. Biebel, S. Bludman, W. Buchm\"uller, M. Kachelriess, M. Kowalski, K. Mannheim, H. Meyer, 
W. Ochs, K. Petrovay, D. Semikoz, and G. Sigl for useful 
discussions. Furthermore, I thank the OPAL collaboration for their unpublished
results on Z decays.
The work of Z. Fodor and S. Katz was partially supported by Hungarian Science Foundation
grants No. OTKA-\-T34980/\-T29803/\-T22929/\-M28413/\-OMFB-1548/\-OM-MU-708. 
\end{acknowledgements}


\begin{thebibliography}{99}


\bibitem[Abe et al.(1999)]{Abe99}
Abe, K. {\em et al.}, hep-ex/9908033
%%CITATION = HEP-EX 9908033;%% 
\bibitem[Abe et al.(unpublished)]{Abe01}
Abe, K. {\em et al.}, OPAL PN299, unpublished.
\bibitem[Ab\-reu et al.(1995)]{Abreu95}
Abreu, P. {\em et al.}, Nucl. Phys. B 444, 3,  1995.
%%CITATION = NUPHA,B444,3;%%
\bibitem[A\-kers et al.(1994)]{Akers94} 
Akers, R. {\em et al.}, Z. Phys. C 63, 181,  1994.
%%CITATION = ZEPYA,C63,181;%%
%\cite{Arnaud:2001gt}
\bibitem[Arnaud et al. (2001)]{Arnaud:2001gt}
%N.~Arnaud, M.~Barsuglia, M.~A.~Bizouard, F.~Cavalier, M.~Davier, P.~Hello and T.~Pradier,
Arnaud, N. {\em et al.},
%``Gravity wave and neutrino bursts from stellar collapse: A sensitive  test of neutrino masses,''
%arXiv:
hep-ph/0109027.
%%CITATION = HEP-PH 0109027;%%
\bibitem[Ave et al.(2000)]{Ave00}
Ave, M. {\em et al.}, Phys. Rev. Lett. 85, 2244, 2000.
%%CITATION = ASTRO-PH 0007386;%%
\bibitem[Bah\-call and Wax\-man(2000)]{BW99} 
Bahcall, J.N. and Waxman, E., 
Astrophys. J. 542, 543, 2000.
%%CITATION = HEP-PH 9912326;%%
%\cite{Bahcall:2001yr}
\bibitem[Bahcall and Waxman (2001)]{Bahcall:2001yr}
Bahcall, J.N. and Waxman, E., 
%``High energy astrophysical neutrinos: The upper bound is robust,''
Phys.\ Rev.\ D {64} (2001) 023002.
%[arXiv:hep-ph/9902383].
%%CITATION = HEP-PH 9902383;%%
\bibitem[Baltrusaitis et al.(1985)]{Baltrusaitis85} 
Baltrusaitis, R.M. {\em et al.},
Phys. Rev. D 31, 2192, 1985.
%%CITATION = PHRVA,D31,2192;%%
\bibitem[Barwick (unpublished)]{Barwick00} 
Barwick, S., 
%talk at RADHEP 2000,
%UCLA workshop,
www.ps.uci.edu/\~{}amanda
%\cite{Beresinsky:1969qj}
\bibitem[Berezinsky and Zatsepin (1969)]{Beresinsky:1969qj}
Berezinsky, V.S. and Zatsepin, G.T., 
%``Cosmic Rays At Ultrahigh-Energies (Neutrino?),''
Phys.\ Lett.\ B {28} (1969) 423.
%%CITATION = PHLTA,B28,423;%%
%
%\cite{Beresinsky:1970}
\bibitem[Berezinsky and Zatsepin (1970)]{Beresinsky:1970}
Berezinsky, V.S. and Zatsepin, G.T.,
Sov.\ J.\ Nucl.\ Phys.\  {11} (1970) 111
[Yad.\ Fiz.\ {11} (1970) 200].
%%CITATION = SJNCA,11,111;%%
%\cite{Berezinsky:1979pd}
\bibitem[Berezinsky (1979)]{Berezinsky:1979pd}
Berezinsky, V.S., 
%``High-Energy Neutrino Astronomy Versus Gamma Astronomy. (Talk),''
in {\em Proc. DUMAND Summer Workshops, Learned, J.G. (Ed.),  
Khabarovsk and Lake Baikal, 22-31 Aug 1979, Hawaii DUMAND Center, 
University of Hawaii, 1980, pp. 245-261}.
\bibitem[Bertou et al.(2000)]{Bertou00}
%Bertou, X., Boratav, M., and Letessier-Selvon, A., 
Bertou, X. {\em et al.},
Int. J. Mod. Phys. A15, 2181, 2000.
%%CITATION = ASTRO-PH 0001516;%%
\bibitem[Bhattacharjee and Sigl(2000)]{BS00} 
Bhattacharjee, P. and Sigl, G., Phys. Rept. 327, 109, 2000.
%%CITATION = ASTRO-PH 9811011;%%
\bibitem[Biebel (priv. comm.)]{Biebel01} 
Biebel, O., private communication.
\bibitem[Bird et al.(1993)]{Bird93} 
Bird, D.J. {\em et al.}, Phys. Rev. Lett. 71, 3401, 1993.
%%CITATION = PRLTA,71,3401;%%
\bibitem[Bird et al.(1994)]{Bird94} 
Bird, D.J. {\em et al.}, Astrophys. J. 424, 491, 1994.
%%CITATION = ASJOA,424,491;%%
\bibitem[Bird et al.(1995)]{Bird95} 
Bird, D.J. {\em et al.}, Astrophys. J. 441, 144, 1995.
%%CITATION = ASJOA,441,144;%%
\bibitem[Bla\-n\-co-Pillado et al.(2000)]{BVZ00}
% Blanco-Pillado, J.J., V\'azquez, R.A., and Zas, E.,  
Blanco-Pillado, J.J. {\em et al.}
Phys. Rev. D 61, 123003, 2000.
%%CITATION = ASTRO-PH 9902266;%%
\bibitem[Buskulic et al.(1995)]{Buskulic95}
Buskulic, D. {\em et al.}, Z. Phys. C 66,  355, 1995.
%%CITATION = ZEPYA,C66,355;%%
\bibitem[Capelle et al.(1998)]{Capelle98} 
Capelle, K.S. {\em et al.},  
%Cronin, J.W., Parente, G., and Zas, E., 
Astropart. Phys. 8, 321, 1998.
%%CITATION = ASTRO-PH 9801313;%%
%\cite{Czakon:2001uh}
\bibitem[Czakon et al. (2001)]{Czakon:2001uh}
%M.~Czakon, J.~Gluza, J.~Studnik and M.~Zralek,
Czakon, M. {\em et al.},
%``In quest of neutrino masses at O(eV) scale,''
%arXiv:
hep-ph/0110166.
%%CITATION = HEP-PH 0110166;%%
%\cite{daCosta:1996nt}
\bibitem[da Costa et al. (1996)]{daCosta:1996nt}
da Costa, L.N. {\em et al.}, 
%W.~Freudling, G.~Wegner, R.~Giovanelli, M.~P.~Haynes and J.~J.~Salzer,
%``The Mass Distribution in the Nearby Universe,''
Astrophys. J. 468, L5, 1996.
%arXiv:astro-ph/9606144.
%%CITATION = ASTRO-PH 9606144;%%
\bibitem[De\-kel et al.(1999)]{Dekel99} 
Dekel, A. {\em et al.},
Astrophys. J. 522, 1,  1999.
%%CITATION = ASJOA,522,1;%%
\bibitem[Efimov et al.(1991)]{Efimov91} 
Efimov, N.N. {\em et al.}, in
{\em Proc. of the Astrophysical Aspects of the Most Energetic Cosmic
Rays} (World Scientific, Singapore, 1991).
%\cite{Engel:2001hd}
\bibitem[Engel and Stanev (2001)]{Engel:2001hd}
Engel, R. and Stanev, T. 
%``Neutrinos from propagation of ultra-high energy protons,''
Phys.\ Rev.\ D {64} (2001) 093010.
%[arXiv:astro-ph/0101216].
%%CITATION = ASTRO-PH 0101216;%%
\bibitem[Fargion et al.(1999)]{FMS99}
Fargion, D., Mele, B. and Salis, A., Astrophys. J. 517, 725, 1999.
%%CITATION = ASTRO-PH 9710029;%%
\bibitem[Fargion et al.(2001)]{FGDTK01}
Fargion, D. {\em et al.}, astro-ph/0102426
%%CITATION = ASTRO-PH 0102426;%%
%\cite{Farzan:2001cj}
\bibitem[Farzan et al.(2001)]{Farzan:2001cj}
%Y.~Farzan, O.~L.~Peres and A.~Y.~Smirnov,
%Farzan, Y., Peres, O.L.G., and Smirnov, A.Yu., 
Farzan, Y. {\em et al.}, 
%``Neutrino mass spectrum and future beta decay experiments,''
Nucl.\ Phys.\ B {612} (2001) 59.
%[arXiv:hep-ph/0105105].
%%CITATION = HEP-PH 0105105;%%
\bibitem[Fo\-dor and Katz(2001a)]{FK01a} 
Fodor, Z. and Katz, S.D., Phys. Rev. D 63, 023002,  2001.
%%CITATION = HEP-PH 0007158;%%
\bibitem[Fo\-dor and Katz(2001b)]{FK01b} 
Fodor, Z. and Katz, S.D. Phys. Rev. Lett. 86, 3224, 2001.
%%CITATION = HEP-PH 0008204;%% 
\bibitem[Fo\-dor et al.(2001a)]{FKR01a}
Fodor, Z., Katz, S.D., and Ringwald, A., 
hep-ph/0105064
%%CITATION = HEP-PH 0105064;%%
\bibitem[Fo\-dor et al.(in prep.)]{FKR01b}
Fodor, Z., Katz, S.D., and Ringwald, A., 
in preparation.
%\cite{Gandhi:2000kq}
\bibitem[Gandhi (2000)]{Gandhi:2000kq}
Gandhi, R., 
%``Ultra-high energy neutrinos: A review of theoretical and  phenomenological issues,''
Nucl.\ Phys.\ Proc.\ Suppl.\  {91} (2000) 453.
%[hep-ph/0011176].
%%CITATION = HEP-PH 0011176;%%
\bibitem[Gel\-mi\-ni and Kusenko (1999)]{GK99} 
Gel\-mi\-ni, G. and Kusenko, A.,  Phys. Rev. Lett. 82, 5202, 1999.
%%CITATION = HEP-PH 9902354;%%
%\cite{Gelmini:2000ds}
\bibitem[Gel\-mi\-ni and Kusenko (2000)]{Gelmini:2000ds}
Gelmini, G. and Kusenko, A., 
%``Unstable superheavy relic particles as a source of neutrinos  
%responsible for the ultrahigh-energy cosmic rays,''
Phys.\ Rev.\ Lett.\  {84} (2000) 1378.
%[arXiv:hep-ph/9908276].
%%CITATION = HEP-PH 9908276;%%
\bibitem[Gorham et al.(2001)]{Gorham01} 
Gorham, P.W. {\em et al.}, astro-ph/0102435
%%CITATION = ASTRO-PH 0102435;%%
\bibitem[Grei\-sen(1966)]{G66} 
Greisen, K., Phys. Rev. Lett. 16, 748, 1966.
%%CITATION = PRLTA,16,748;%%
\bibitem[Groom et al.(2000)]{Groom00} 
Groom, D.E. {\em et al.}, Eur. Phys. J. C 15, 1, 2000.
%%CITATION = EPHJA,C15,1;%%
\bibitem[Guerard(1999)]{Guerard99}
Guerard, C.K., Nucl. Phys. Proc. Suppl. 75A, 380, 1999.
%%CITATION = NUPHZ,75A,380;%%
%\cite{Hu:1998mj}
\bibitem[Hu et al. (1998)]{Hu:1998mj}
%W.~Hu, D.~J.~Eisenstein and M.~Tegmark,
Hu, W. {\em et al.},
%``Weighing neutrinos with galaxy surveys,''
Phys.\ Rev.\ Lett.\  {80} (1998) 5255.
%[arXiv:astro-ph/9712057].
%%CITATION = ASTRO-PH 9712057;%%
\bibitem[Ki\-eda et al.(2000)]{Kieda00} 
Kieda, D. {\em et al.}, in 
{\em Proceedings of the 26th International Cosmic Ray Conference, Salt Lake, 
1999}.
% (AIP, New York, 2000)
%\cite{Kneller:2001cd}
\bibitem[Kneller et al. (2001)]{Kneller:2001cd}
%J.~P.~Kneller, R.~J.~Scherrer, G.~Steigman and T.~P.~Walker,
Kneller, J.P. {\em et al.},
%``When Does CMB + BBN = New Physics?,''
Phys.\ Rev.\ D {64} (2001) 123506.
%[arXiv:astro-ph/0101386].
%%CITATION = ASTRO-PH 0101386;%%
%\cite{Learned:2000sw}
\bibitem[Learned and Mannheim (2000)]{Learned:2000sw}
Learned, J.G. and Mannheim, K., 
%``High-energy neutrino astrophysics,''
Ann.\ Rev.\ Nucl.\ Part.\ Sci.\  {50} (2000) 679.
%%CITATION = ARNUA,50,679;%%
\bibitem[Lawrence et al.(1991)]{Lawrence91} 
%Lawrence, M.A., Reid, R.J.O., and Watson, A.A., 
Lawrence, M.A. {\em et al.},
J. Phys. G 17, 773, 1991.
%%CITATION = JPHGB,G17,733;%%
%\cite{Ma:1998aw}
\bibitem[Ma (1999)]{Ma:1998aw}
Ma, C.P., 
%``Neutrinos and dark matter,''
%arXiv:
astro-ph/9904001.
%%CITATION = ASTRO-PH 9904001;%%
%\cite{Mannheim:2001wp}
\bibitem[Mannheim et al. (2001)]{Mannheim:2001wp}
%Mannheim, K.,  Protheroe, R.J., and Rachen, J.P., 
Mannheim, K. {\em et al.},
%``On the cosmic ray bound for models of extragalactic neutrino  production,''
Phys.\ Rev.\ D {63} (2001) 023003.
%[arXiv:astro-ph/9812398].
%%CITATION = ASTRO-PH 9812398;%%
%\cite{McKellar:2001hk}
\bibitem[McKellar et al. (2001)]{McKellar:2001hk}
McKellar, B.H. {\em et al.}, 
%M.~Garbutt, G.~J.~Stephenson and T.~Goldman,
%``Neutrino clustering and the Z burst model,''
%arXiv:
hep-ph/0106123.
%%CITATION = HEP-PH 0106123;%%
\bibitem[Ormes et al.(1997)]{Ormes97}
Ormes, J.F. {\em et al.}, in {\em Proc. of the 25th International Cosmic 
Ray Conference, Potchefstroom, 1997}. 
%(Potchefstroom University, Potchefstroom, 1997).
%\cite{Osland:2001pm}
\bibitem[Osland and Vigdel (2001)]{Osland:2001pm}
Osland, P. and Vigdel, G.,
%``Neutrinoless double beta decay, solar neutrinos and mass scales,''
Phys.\ Lett.\ B {520} (2001) 143
%[arXiv:hep-ph/0107161].
%%CITATION = HEP-PH 0107161;%%
%\cite{Pas:2001nd}
\bibitem[P\"as and Weiler(2001)]{Pas:2001nd}
P\"as, H. and Weiler, T.J., 
%``Absolute neutrino mass determination,''
Phys.\ Rev.\ D {63} (2001) 113015.
%[arXiv:hep-ph/0101091].
%%CITATION = HEP-PH 0101091;%%
%\cite{Primack:2000iq}
\bibitem[Primack and Gross (2000)]{Primack:2000iq}
Primack, J.R. and Gross, M.A., 
%``Hot dark matter in cosmology,''
%arXiv:
astro-ph/0007165.
%%CITATION = ASTRO-PH 0007165;%%
\bibitem[Primack (priv. comm.)]{Primack:2001}
Primack, J.R., private communication.
%\cite{Protheroe:1996ft}
\bibitem[Pro\-the\-roe and Johnson (1996)]{Protheroe:1996ft}
Protheroe, R.J. and Johnson, P.A., 
%``Propagation of ultrahigh-energy protons over cosmological distances and implications for topological defect models,''
Astropart.\ Phys.\  {4} (1996) 253
[Erratum-ibid.\  {5} (1996) 215].
%[astro-ph/9506119].
%%CITATION = ASTRO-PH 9506119;%%
%\cite{Protheroe:1999ei}
\bibitem[Pro\-the\-roe (1999)]{Protheroe:1999ei}
Protheroe, R.~J., 
%``High energy neutrino astrophysics,''
Nucl.\ Phys.\ Proc.\ Suppl.\  {77} (1999) 465.
%%CITATION = NUPHZ,77,465;%%
\bibitem[Rou\-let(1993)]{R93} 
Roulet, E., Phys. Rev. D 47, 5247, 1993.
%%CITATION = PHRVA,D47,5247;%%
\bibitem[Seckel et al. (2001)]{Seckel:2001}
Seckel, D. {\em et al.},
in {\em Proc. 27th International Cosmic Ray Conference, Hamburg, 2001, pp. 1137-1140}. 
%\cite{Sigl:priv}
\bibitem[Sigl (priv. comm.)]{Sigl:priv}
Sigl, G., private communication
\bibitem[Sreekumar et al. (1998)]{Sreekumar}
Sreekumar, P. {\em et al.}, 
Astrophys. J. {494}, 523 (1998).
\bibitem[Takeda et al.(1998)]{Takeda98} 
Takeda, M. {\em et al.}, Phys. Rev. Lett. 81, 1163, 1998.
%%CITATION = ASTRO-PH 9807193;%%
\bibitem[Takeda et al.(1999)]{Takeda99} 
Takeda, M. {\em et al.}, 
astro-ph/9902239
%%CITATION = ASTRO-PH 9902239;%%
%\cite{Tinyakov:2001nr}
\bibitem[Tinyakov and Tkachev (2001)]{Tinyakov:2001nr}
Tinyakov, P.G. and Tkachev, I.I., 
%``BL Lacertae are sources of the observed ultra-high energy cosmic rays,''
%arXiv:
astro-ph/0102476.
%%CITATION = ASTRO-PH 0102476;%%
\bibitem[Wang et al.(2001)]{WTZ01}
Wang, X., Tegmark, M., and Zaldarriaga, M., 
astro-ph/0105091.
%%CITATION = ASTRO-PH 0105091;%%
\bibitem[Waxman(1995)]{W95} 
Waxman, E., Astrophys. J. 452, L1, 1995. 
%%CITATION = ASTRO-PH 9508037;%%
\bibitem[Waxman(1998)]{W98} 
Waxman, E., astro-ph/9804023.
%%CITATION = ASTRO-PH 9804023;%%
%\cite{Waxman:1999yy}
\bibitem[Waxman and Bahcall (1999)]{Waxman:1999yy}
Waxman, E. and Bahcall, J.N., 
%``High energy neutrinos from astrophysical sources: An upper bound,''
Phys.\ Rev.\ D {59} (1999) 023002. 
%[arXiv:hep-ph/9807282]. 
%%CITATION = HEP-PH 9807282;%%
\bibitem[Weiler(1982)]{W82} 
Weiler, T., Phys. Rev. Lett. 49, 234, 1982.
%%CITATION = PRLTA,49,234;%% 
\bibitem[Weiler(1999)]{W99}
Weiler, T.J., Astropart. Phys. 11, 303, 1999;
12, 379, 2000 (Err.).
%%CITATION = HEP-PH 9710431;%%
%\cite{Vissani:2001ci}
\bibitem[Vissani (2001)]{Vissani:2001ci}
Vissani, F., 
%``Non-oscillation searches of neutrino mass in the age of oscillations,''
Nucl.\ Phys.\ Proc.\ Suppl.\  {100} (2001) 273.
%[arXiv:hep-ph/0012018].
%%CITATION = HEP-PH 0012018;%%
\bibitem[Yoshida and Teshima (1993)]{Yoshida:1993pt}
Yoshida, S. and Teshima, M. 
%``Energy spectrum of ultrahigh-energy cosmic rays with extragalactic origin,''
Prog.\ Theor.\ Phys.\  {89} (1993) 833.
%%CITATION = PTPKA,89,833;%%
%\cite{Yoshida:1997ie}
\bibitem[Yoshida et al. (1997)]{Yoshida:1997ie}
%Yoshida, S., Dai, H., Jui, C.C., and Sommers, P.,
Yoshida, S. {\em et al.},
%``Extremely high energy neutrinos and their detection,''
Astrophys.\ J.\  {479} (1997) 547.
%[astro-ph/9608186].
%%CITATION = ASTRO-PH 9608186;%%
\bibitem[Yoshida et al.(1998)]{Y99} 
Yoshida, S., Sigl, G., and Lee, S.,
Phys. Rev. Lett. 81, 5505, 1998.
%%CITATION = HEP-PH 9808324;%% 
\bibitem[Yoshida et al. (2001)]{Yoshida:2001}
Yoshida, S. {\em et al.}, 
in {\em Proc. 27th International Cosmic Ray Conference, Hamburg, 2001, pp. 1142-1145}. 
\bibitem[Zatsepin and Kuzmin(1966)]{ZK66}
Zatsepin, G.T. and Kuzmin, V.A., Pisma Zh. Eksp. Teor. Fiz. 4, 114, 1966.
%%CITATION = JTPLA,4,78;%%

\end{thebibliography}
\end{document}